# Quantum-enhanced reinforcement learning for finite-episode games with discrete state spaces

Florian Neukart[*1], David Von Dollen[1], Christian Seidel[2], and Gabriele Compostella[2]

[1]Volkswagen Group of America
[2]Volkswagen Data:Lab

## Abstract

Quantum annealing algorithms belong to the class of metaheuristic tools, applicable for solving binary optimization problems. Hardware implementations of quantum annealing, such as the quantum annealing machines produced by D-Wave Systems [1], have been subject to multiple analyses in research, with the aim of characterizing the technology's usefulness for optimization and sampling tasks [2–16]. Here, we present a way to partially embed both Monte Carlo policy iteration for finding an optimal policy on random observations, as well as how to embed $n$ sub-optimal state-value functions for approximating an improved state-value function given a policy for finite horizon games with discrete state spaces on a D-Wave 2000Q quantum processing unit (QPU). We explain how both problems can be expressed as a quadratic unconstrained binary optimization (QUBO) problem, and show that quantum-enhanced Monte Carlo policy evaluation allows for finding equivalent or better state-value functions for a given policy with the same number episodes compared to a purely classical Monte Carlo algorithm. Additionally, we describe a quantum-classical policy learning algorithm. Our first and foremost aim is to explain how to represent and solve parts of these problems with the help of the QPU, and not to prove supremacy over every existing classical policy evaluation algorithm.

[*]Corresponding author: `florian.neukart@vw.com`

# 1 Introduction

The physical implementation of quantum annealing that is used by the D-Wave machine minimizes the two-dimensional Ising Hamiltonian, defined by the operator $H$:

$$H(s) = \sum_{i \in V} h_i s_i + \sum_{ij \in E} J_{ij} s_i s_j. \qquad (1)$$

Here, $s$ is a vector of $n$ spins $\{-1, 1\}$, described by an undirected weighted graph with vertices ($V$) and edges ($E$). Each spin $s_i$ is a vertex (in $V$), $h_i$ represents the weights for each spin, and $J_{ij}$ are the strengths of couplings between spins (edges in $E$). Finding the minimum configuration of spins for such a Hamiltonian is known to be NP-hard. The QPU is designed to solve quadratic unconstrained binary optimization (QUBO) problems, where each qubit represents a variable, and couplers between qubits represent the costs associated with qubit pairs. The QPU is a physical implementation of an undirected graph with qubits as vertices and couplers as edges between them. The functional form of the QUBO that the QPU is designed to minimize is:

$$\text{Obj}(x, Q) = x^T \cdot Q \cdot x, \qquad (2)$$

where $x$ is a vector of binary variables of size $N$, and $Q$ is an $N \times N$ real-valued matrix describing the relationship between the variables. Given the matrix $Q$, finding binary variable assignments to minimize the objective function in Equation 2 is equivalent to minimizing an Ising model, a known NP-hard problem [16, 17].

To directly submit a problem to a D-Wave QPU, the problem must be formulated as either an Ising model or a QUBO instance. The spins are represented by superconducting loops of niobium metal, which are the quantum bits (qubits). The preparation of states by the quantum annealer is done such that the initial configuration of all spins are purely quantum, in uniform superposition in both available states. The quantum annealing system then slowly evolves the system to construct the problem being minimized (described by the QUBO matrix $Q$, or the Ising model's $h$ and $J$).

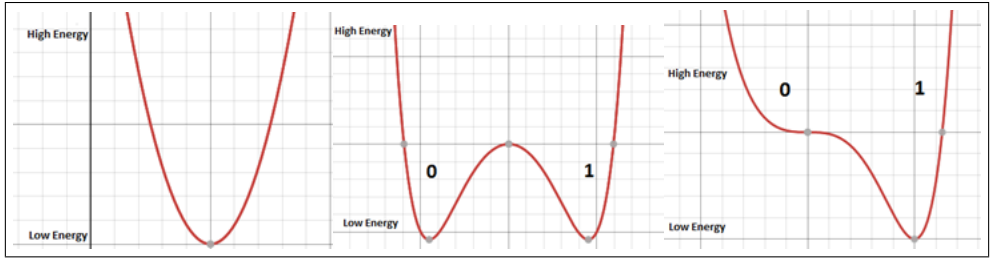

Figure 1: Left: Initial superposition represented by the minimum energy of the 1-qubit system. Middle: Double-well potential caused by quantum annealing. Right: Tilting of the double-well potential by the application of external magnetic fields.

During the annealing, the qubits transition from being quantum objects to classical objects, whose states can then be readout without destroying quantum information. Due to superconductivity, superposition is maintained by inducing a current in both directions at once, corresponding to the 1 and 0 states. Concerning an energy function, the superposition state corresponds to the lowest point in the function's single valley (Fig. 1). During the process of quantum annealing, a barrier is raised, turning the energy diagram into a double-well potential, where the low point of the left valley represents the state 0, and the low point on the right the state 1. The initial probability of being in either the state 0 or the state 1, is given an equally-weighted probability, $\frac{1}{2}$ for each.

By construction, the qubits always start in the minimum energy configuration. As the problem is introduced, other energy levels get closer to the lowest energy level, which poses a challenge: the smaller the gap between the energy states, the more likely it is that the system will jump from its lowest energy configuration into one of the excited states. The so-called minimum gap is the point where the first excited state approaches the ground state closely, and thermal fluctuations or too short annealing times could result in the system jumping from the ground state to an excited state. If the annealing happens slow enough, meaning that the system stays in the ground state, it follows an adiabatic process. This is particularly worth mentioning, because the larger the system is, the more unlikely it is to remain in the ground state during the annealing cycle.

However, even if the quantum system settles in one of the excited states, this result can be useful. As the system size grows, it becomes harder to classically validate the optimal (minimum energy) solution. Brute force is, for example, classically intractable for even a modest number of qubits ($\sim 10^4$). If an excited state is still low-energy enough to be useful in a practical setting, much computation time can be saved by using the quantum annealing system.

Additional built-in devices in D-Wave quantum annealers are the couplers, which allow multi-qubit entanglement. Entanglement refers to correlating qubits in a way that they cannot be described as separate subsystems and act as a single quantum object. Thus, considering a two qubit system, the change of one qubits state also affects the second qubit, and they can be correlated such that if $q_1$ ends up in a certain state, $q_2$ is forced to take the same or opposite state. However, the object $q_1 q_2$ can take 4 different states $S = 00, 01, 10, 11$, and the relative energy of these states depends on the biases of each qubit, and the coupling between them. This is how a program on the D-Wave quantum annealing system is defined [18]: choosing a set of biases and couplings defines an energy-landscape, whose global minima correspond to the problem being solved. At the end of the quantum algorithm, the system ends up in the minimum energy configuration, thus solving the problem. A classical algorithm would represent the problem in the same way, but where a classical algorithm can walk the surface, the qubits are capable of using *quantum tunneling* to pass through the energy barriers of the surface. Once entangled, qubits can tunnel together through the energy barriers from one configuration to another. Previous publications have shown how quantum effects in the D-Wave QPU, such as entanglement, superposition, and tunneling, help the QPU solve combinatorial optimization problems [19].

# 2 Formalization of the reinforcement learning problem using the Markov decision process framework

The reinforcement learning problem can be described by the Markov decision process (MDP), which consists of states, actions that can be taken in each of these states, state transitions, and a reward function. Furthermore, a distinction between discrete and continuous state spaces, and finite vs. infinite horizon episodes have to be made [20, 21].

## 2.1 States

A set of states $S$ is defined as finite set $\{s^1, ..., s^N\}$, and each $s \in S$ is described by its features, and taking a self-driving cars (SDC) as example, some of the features are the SDC's position in the world, other traffic participants (position, velocity, trajectory,...), infrastructure (traffic lights, road condition, buildings, construction zones,...), weather conditions, ..., thus, everything that matters for the SDC in a certain situation at a certain time. Some states are legal, some are illegal, which is based upon the combination of features. It is, for example, illegal for the SDC to occupy space that's occupied by a building or another vehicle at the same time.

## 2.2 Actions

In any given state an agent should be able to evaluate and execute a set of actions, which is defined as the finite set $\{a^1, ..., a^K\}$, where $K$ characterizes the size of the action space $|A| = K$. Before we mentioned illegal states, which may result from applying an illegal action for a state, so not every action may be applied in every given state. Generally, and we describe the set of actions that may be applied in a particular state $s \in S$ by $A(s)$, where $A(s) \subseteq A$ and $A(s) = A$ for all $s \in S$. As we also need to account for illegal states, this can be modeled by a precondition function:

$$f_{legal} : S \times A \to \{true, false\}, \qquad (3)$$

## 2.3 Transition

A transition from a state $s \in S$ to a consecutive state $s' \in S$ occurs after the agent executes an action in the former. The transition is usually not encoded in hard rules, but given by a probability distribution over the set of possible transitions (different actions will result in different states). This is encoded into a transition function, which is defined as

$$T : S \times A \times S \to [0, 1], \qquad (4)$$

which states that the probability of transitioning to state $s' \in S$ after executing $A(s)$ in $s \in S$ is $T(s, a, s')$. For all actions $a$, all states $s$, and consecutive states after transition $s'$, $T(s, a, s') \geq 0$ and $T(s, a, s') \leq 1$, and for all actions $a$, $\sum_{s'} T(s, a, s')=1$. Given that the result of an action does not depend on the history (previous states and actions), and depends only on the current state, then it is

called Markovian:

$$P(s_{t+1}|s_t, a_t, s_{t-1}, a_{t-1}) = P(s_{t+1}|s_t, a_t) = T(s_t, a_t, s_{t+1}), \quad (5)$$

where $t = 1, 2, ...$ are the time steps. In such a system, only the current state $s$ encodes all the information required to make an optimal decision.

## 2.4 Reward

The reward is given by a function for executing an action in a given state, or for being in a given state. Different actions in a state $s$ may result in different rewards, and the respective reward function is either defined as

$$R : S \to \mathbb{R}, \quad (6)$$

or as

$$R : S \times A \to \mathbb{R}, \quad (7)$$

or as

$$R : S \times A \times S \to \mathbb{R}, \quad (8)$$

Equation 6 gives the reward obtained in states, Equation 7 gives rewards for performing actions in certain states, and Equation 8 gives rewards transitions between states.

# 3 Finite horizon, discrete state space

We have been focusing on finite horizon and discrete state space games so far, and in games such as the one used to describe the introduced quantum reinforcement learning example, the foundation for finding "good states" and subsequently approximating an optimal policy are observations in the form of completed games (episodes). In Black Jack, a state is given by the player's current sum (12-21), the dealer's one showing card (given by the values 1-10, where 1 is an ace), and whether or not the player holds a usable ace (given by 0 or 1, where 0 is no usable ace). The two actions the player can execute are either to stick (=stop receiving cards), to hit (=receive another card), given by $a \in \{0, 1\}$. Every episode is composed of multiple state-action pairs, and for each state-action pair a reward value is given. Although the same state-action pairs may occur in different episodes, it is very unlikely that the reward for these in two episodes is the same, as the reward in a state not only depends from the state, but also from the previous states and actions. What we intend to approximate from $n$ complete episodes is the optimal policy under consideration of all states.

# 4 Difference Monte Carlo and quantum-enhanced Monte-Carlo policy evaluation

Monte Carlo methods learn from complete sample returns, which implies that they are defined for episodic tasks only. An update happens after each episode, so learning happens directly from experience. The goal is to learn the optimal policy given some episodes, and the basic idea is to average the returns after a state $s$ was visited. A distinction has to be made about whether the returns for each visit to $s$ in an episode are averaged, or only first time visits. Each of the approaches converges asymptotically. As for the used game each state occurs only once per episode, first-visit Monte Carlo policy evaluation is described in Algorithm 1:

---
**Algorithm 1** Monte Carlo policy evaluation
---
**Initialize:** $\pi, V, Return(s)$
**Repeat forever:**
(a) Generate an episode using $\pi$
(b) For each state $s$ in current episode
    $R \leftarrow$ return following the first occurrence of $s$
    Append $R$ to $Return(s)$
    $V(s) \leftarrow$ average $(Return(s))$
---

where $\pi$ is the policy to be evaluated, $V$ a state-value function, and $Return(s)$ is initialized as empty list and will hold all $s \in S$. In the Black Jack example, our aim is to have a card sum that is greater than the dealer's sum, but in the same instance we must not exceed 21. A reward of +1 is given for winning the game, 0 reward for a draw game, and -1 for losing the game. The state-value function we approximate is based on the policy $\alpha$, which is defined as "stick if the sum is 20 or 21, and else hit". For this, we simulated many Black Jack games using policy $\alpha$ and averaged the returns following each state. Finding the state-value function with a purely classical Monte Carlo algorithm requires thousands of observations (simulations), based on which the returns following each state are averaged (see Figs. 2, 3, showing the "no usable ace"-scenario only).

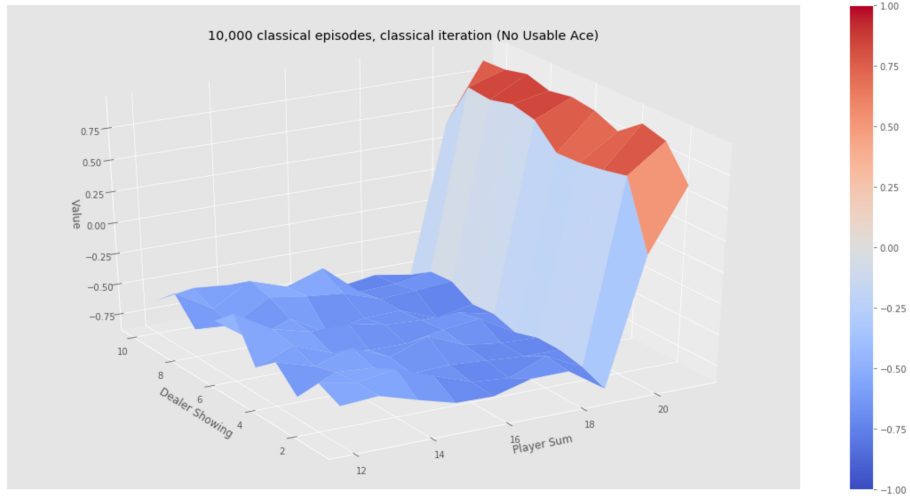

Figure 2: Monte Carlo: 10,000 episodes, policy $\alpha$, no usable ace

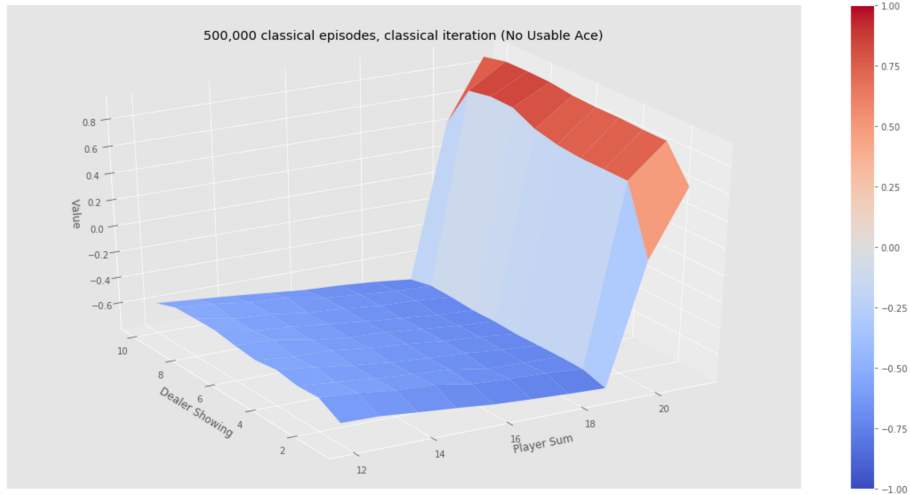

Figure 3: Monte Carlo: 500,000 episodes, policy $\alpha$, no usable ace

Additionally, we simulated many blackjack games using random policy $\beta$, which would choose a random action in each state, averaged the returns following each state, and used this to find a policy only on observation. The respective value function is plotted in 4.

## 4.1 State-value function approximation

In the first example, we generate $n$ sub-optimal state-value functions by purely classical Monte Carlo policy evaluation. We want to emphasize that the formulation of the algorithm is such that either $n$ classically generated (sub-optimal) state-value functions can be embedded for approximating an improved state-value function, or the state-value functions can be generated quantum-enhanced by supplying episodes with their respective (non-averaged) rewards to the algorithm. Both may also be combined in a nested algorithm. The QUBO formulation is such that it will find the preferable state-action pairs based on their rewards over the episodes, and once this has been completed for $m$ episodes, $n$ state-value functions can be generated according to the same proceeding:

- Based on ( episodes, find $n$ sub-optimal state-value functions classically and embed each state-action pair with the respective value directly on the QPU and so help to approximate the optimal state-value function, which is better than each of the $n$ sub-optimal state-value functions. How identical state-action pairs over different state-value functions are dealt with will be explained below.

- Based on ( episodes, we embed state-action pairs directly on the QPU and approximate a policy directly from the observations.

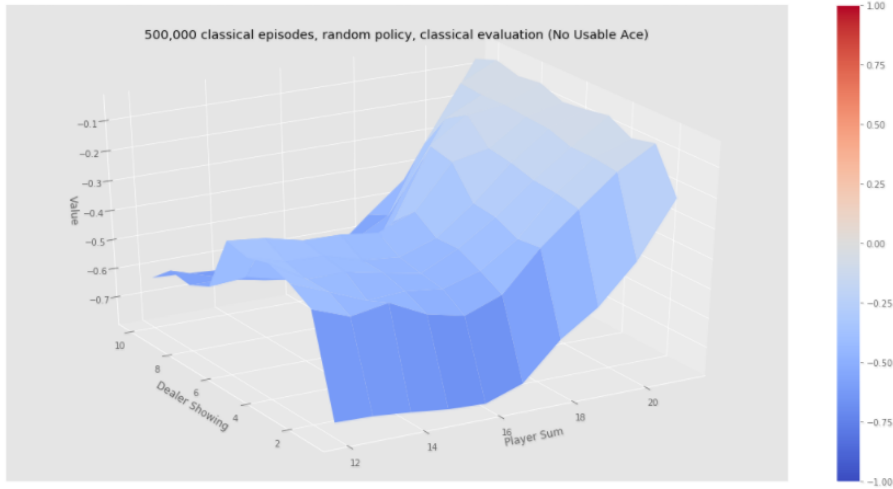

Figure 4: Monte Carlo: 500,000 episodes, policy $\beta$, no usable ace

When using the QPU to approximate the state-value function, we first classically simulated $m$ Black Jack games according to the policy $\alpha$. We did this $n$ times, and generated $n$ sub-optimal state-value functions. We then embedded these state-value functions on the QPU such that the QPU acts as a filter, which identifies the state-action pairs that do not positively contribute to approximating the optimal state-value function, and removes these from the result. Still, the result may contain identical state-action pairs with different rewards, but those are averaged over. In our experiments, with the help of the QPU we were on average able to filter out approximately 1/2 of the state-action-value triples from the initially supplied observations. Due to the partially stochastic nature of the game and the varying episode-length, the number of different state-action-value triples may vary.

## 4.2 Policy iteration

Concerning policy iteration, we chose random actions in each state, observed the obtained reward, and use this to learn a policy. The proceeding is similar to the latter example, except that we we do not start with a given policy, such as *alpha* described above, but try to find a policy (what to do in each state) based observations. By applying the QPU-filter to the original data, we show that we are able to learn policies of similar quality as with the purely classical algorithm, although the number of states in the original data can be reduced by up to 2/3.

## 5 Formulation of the problem as QUBO

The following explanations are based on classically generated, sub-optimal state-value functions, which are embedded on the QPU. The resulting state-value function is compared with single sub-optimal state-value functions, as well as with a state-value function which was obtained by averaging over all sub-optimal state-value functions. The sub-optimal state-value functions in this example were generated using 100 - 500 ($m$) episodes each. We varied the number of generated state-value functions $n$ from 2 - 7. Unsurprisingly, we obtained the best results with 500 episodes and 7 sub-optimal state-value functions. Increasing $n$ by 1 and/ or $m$ by 50 did not allow for embedding the problem on the QPU without splitting it into sub-problems. Here, we explain how to use qubits for representing state-action pairs, and how to determine the entries of the QUBO by using the value per state-action pair we obtained in our sub-optimal state-value functions. In a nutshell, we use the 2-dimensional physical architecture of the chip as if it was 3-dimensional by representing the problem as a tensor-like structure, but the introduced method generalizes to more complex problems with $n$ dimensions. Based on the following explanations it will become obvious that the more observations are supplied, the more qubit and connections between single qubits are required. Thus, as we have to deal with sparse connectivity and a limited number of qubits, it is in our interest to find a formulation of the problem that allows us to find a sufficiently good state-value function for a given policy by keeping the number of observations small.

### 5.1 Conditions

What we approximate is a state-value function for a policy based on observations (complete episodes), and the reward that is given for an action in a given state. Certainly, it is not possible to just rely on one observation, as in each episode we can see a certain state-action pair only once, and thus only are given one reward-value. It may be that in one episode a certain action for a certain state is rewarded higher than in another episode, depending on which actions have been executed in which states in the past. Furthermore, a game such as Black Jack features stochastic components, thus we assume it is possible to make some assumptions about future states (i.e. the player's current sum in an ongoing episode will not decrease in the next state), but it cannot be accurately predicted. This means that we should not ignore the future states in a given state completely, but have to include them into our considerations to a certain degree. Therefore, for formulating the problem as a QUBO we need to consider the following conditions:

- As identical state-action pairs may be rewarded differently in different episodes, we must assume that in the worst case the execution of a certain action given a certain state is positively rewarded in one episode, and negatively rewarded in another episode. As we intend to not only identify random successful state-action pairs, but successful sequences (see condition 2), averaging or any other means of aggregating the rewards of state-action pairs is prohibited, and the QUBO must be formulated such identical state action-pairs over the episodes are given different entries.

- As the reward in a certain state partly depends on what happened in the past, and we are dealing with multiple episodes and want to consider the success of an action in a given state by evaluating the future states, the QUBO must be formulated such that both actions taken in past and future states are taken into account, which allows us to not only determine preferable state-action pairs, but preferable state-action chains by indirectly encoding the statistical probability of consecutive state-action pairs being successful or not while approaching the end of the game. If, in our observed episodes, we find state-action-sequences $(s', a', s'', a'', s''', a'''...)$ and $(s*, a*, s**, a**, s***, a***...)$, and the latter comes with higher rewards for consecutive state-action pairs, we assume that this is a more successful sequence. If $k >> l$ and a sequence $b$ gives positive reward $k$ times, and negative reward $l$ times, $b$ will be considered as successful sequence, as $k$ times the cumulative positive reward outweighs the $l$ times cumulative negative reward. If for $b$ only $k > l$, other successful sequences may outweigh $b$, and it may not contribute to and appear in the optimal policy. If for $b$ $k < l$ or even only $k \approx l$ then it must not be considered in the resulting optimal policy.

## 5.2 QUBO

Each of the $n$ episodes consists of a different number of states $S_n = \{s_1^n, ..., s_m^n\}$, where $m$ is the number of states in a given episode and may vary over the episodes. For each state $s_x^n$ in $S_n$ we may also see varying actions and rewards, depending on the history that lead us to $s_x^n$. Furthermore, we may see different actions, depending on whether the player behaved risk-affine or risk-averse an episode under consideration. Each of the $s_x^n$ must be given a separate entry in the QUBO matrix, as not only the rewards for the state itself, but also the consecutive state-action pairs including their respective rewards may differ. Therefore, the first step is to iterate over all episodes, create a list $L$ of length $l$ containing state-action pairs and the observed negative rewards $r * (-1) * r_f$ in a temporary dictionary, where $r_f$ is a factor applied to scale the reward up or down, depending on the size of the real values. In this given problem, all rewards ranged from (-1) to 1, and we had best successes in scaling up the rewards by a factor of 10. Due to fluctuations during the annealing cycle, the energies of different possible solutions must not be too close to each other, as jumps may happen, i.e. from a lower energy-solution (better) to a higher-energy solution (worse). On the other hand, the higher the energy barriers, the more unlikely tunneling will happen and we may get stuck in a local minimum, so the energy values also must not be too far apart. The negative rewards are needed, as maximizing the reward equals minimizing the negative reward, which can be interpreted as energy minimization problem on the QPU. The basic QUBO-entry for each state-action pair $V_{s,a}$ is thus calculated as shown in Equation 9.

$$L_{s,a} = V_{s,a} = R_{s,a} * (-1) * r_f, \tag{9}$$

The QUBO-matrix is an upper triangular $N \times N$-matrix by $i \in \{0, ..., N-1\}$ and $j \in \{0, ..., N-1\}$ and has rows and columns $\{L_{S,A}, L_{S,A}^T\}$, each entry initialized with 0, and is an where $S$ are all observed states and $A$ the respective actions. $L_{S,A}$ is of length $n * M$, where $n$ is the number of observed episodes and $M = \{m_0, ...m_{n-1}\}$ is the number of states in each episode (which may vary). The entries are given by the following functions (Eqs. 10, 11, 12):

$$QUBO(i,j) = \begin{cases} QUBO(i,j) + (L_{i(v)} + L_{j(v)})^2, & \text{if c1} \\ QUBO(i,j) - ((L_{i(v)} + L_{j(v)})^2), & \text{if c2} \\ QUBO(i,j), & \text{otherwise} \end{cases} \quad (10)$$

where

$$c1 : L_{i(s,a)} = L_{j(s,a)} \text{ and } (L_{i(v)} + L_{j(v)}) > 0 \text{ and } i \neq j \quad (11)$$

and

$$c2 : L_{i(s,a)} = L_{j(s,a)} \text{ and } (L_{i(v)} + L_{j(v)}) < 0 \text{ and } i \neq j \quad (12)$$

where $L_{i(v)}$ gives the $i^{th}$ value $v$ in $L_{s,a,v} \in L_{S,A,V}$, $S = \{s_1, s_2, ..., s_n\}$, $A = a_1, a_2, ..., a_n \{a\}$. This basically means that we lock each state-action pair in $L_{S,A}$, and find the respective duplicate state-action pairs over the remaining entries, which are summed and squared to represent the QUBO entries. For this first QUBO manipulation there are several conditions, however, to be considered for writing an entry:

- An entry is only added if it is the summed and squared values are maximum or minimum per $L_{i(v)}$.

- An entry is only added if the state-action pairs $L_{i(s,a)} + L_j(s,a)$ match.

- We intend to find the best action per state given $n$ observations of respective length $m_n$. As it is most likely that identical state-action pairs appear in different episodes, and as each state-action pair is given a separate QUBO-entry, even if it is identical to one added from another episode, they may not have the same value. We increase or decrease these values quadratically, in order to separate them from one another, which results in a separation or an amplification of identical state-actions based on their values. As we are minimizing energies and therefore multiply the state values by $(-1)$: the smaller $-((L_{i(n)} + L_j)^2)$ the better.

We add the diagonal terms as follows Equation 13:

$$QUBO(i,j) = \begin{cases} QUBO(i,j) + (L_{i(v)} + L_{j(v)}), & \text{if } L_{i(s,a)} = L_{j(s,a)} \text{ and } i = j \\ QUBO(i,j), & \text{otherwise} \end{cases} \quad (13)$$

What follows next is the separation of different state values from one another, as we only want to find the optimal policy, which is the best action in a given state considering the future states. Games like Black Jack have a stochastic component, but nevertheless the history resulting in a state and $n$ observations let us statistically determine what the most likely future states in a given states are.

We penalize identical states with different actions as described in Equation 14.

$$QUBO(i,j) = \begin{cases} QUBO(i,j) + p, & \text{if } L_{i(s)} = L_{j(s)} \text{ and } L_{i(a)} \neq L_{j(a)} \text{ and } i \neq j \\ QUBO(i,j), & \text{otherwise} \end{cases} \quad (14)$$

where $p$ is a penalization constant, which should be set to according to the energy scale.

In order to approximate a good policy not only based on high-valued observations, but chains of consecutive states with cumulative high reward, it is possible to consider chains of length $h$ (the horizon) by manipulating the QUBO-entries of $h$ consecutive states. While iterating over all episodes and states, we identify each state $s'$ following a state-action pair $\{s, a\}$, which may differ from episode to episode. For each of the consecutive states $s'$ we create or manipulate the respective entry according to Eqs. 15 - 17.

$$QUBO(i,j) = \begin{cases} QUBO(i,j) + (L_{i(v)} + L_{j+k(v)})^2, & \text{if c1} \\ QUBO(i,j) - ((L_{i(v)} + L_{j+k(v)})^2), & \text{if c2} \\ QUBO(i,j), & \text{otherwise} \end{cases} \quad (15)$$

where

$$c1 : L_{i-(k-1)(s,a)} = L_{j(s,a)} \text{ and } (L_{i(v)} + L_{j+k(v)}) > 0 \text{ and } i \neq j \quad (16)$$

and

$$c2 : L_{i-(k-1)(s,a)} = L_{j(s,a)} \text{ and } (L_{i(v)} + L_{j+k(v)}) < 0 \text{ and } i \neq j \quad (17)$$

We initialize $k$ with 1 so that in the first iteration $i - (k-1) = i$, and in consecutive iterations $i$ increases with $k$, and we always consider consecutive states. As identical state-action pairs from different episodes received separate entries, the more often a chain is successful, the more often all of its respective values are increased at different $i, j$ in the matrix. Due to the penalization in Equation 14 different actions per state are already separated. The bigger the horizon, the more qubits we need to represent the problem, and the smaller the original problem must be so that it can be embedded without splitting it into sub-problems.

## 6  Experimental results and conclusions

By formulating and embedding the QUBO-matrix on the QPU as specified, we are able to show that:

- Given a policy $\alpha$, which is defined as "stick if the sum is 20 or 21, and else hit", and a limited number of observations, we can use the QPU as a filter to identify the states or sequences of states that do not positively contribute to approximating the optimal state-value function. We can reduce the source state-action-value triples by up to 1/2, and by averaging over the remaining values per state, whereby we may still see duplicate state-action pairs in the result, we can generate an improved state-value function. The resulting state-value function, found by the quantum-enhanced Monte Carlo algorithm, is at least equivalent or even better compared to one learned with purely classical Monte Carlo policy evaluation on all given state-action-value triples. Due to the partially stochastic nature of the game and the varying episode-length, the number of different source-state-action-value triples may vary, but here are two examples in support of our explanations: by setting $n = 7$, $m = 500$, we obtained 1072 different state-action-value triples in the source data, with 1004 of them being distinct. With the QPU-filter, we could reduce the number of state-action-value triples to 618, over which we averaged. In another example, with unchanged $n$ and $m$, we ran the algorithm 5 times, and summarized the result to 5393 state-action-value triples in the source data, with 4699 of them being distinct. With the QPU-filter, we could reduce the number state-action-value triples to 3019, over which we averaged. For the latter case, plotted in Figs. 6 and 7, the Euclidean distance from the state-value function found with the quantum-enhanced algorithm to the optimal state-value function found with the purely classical algorithm and 500,000 episodes (with varying numbers of states) is 2.01, whereas the distance from a state-value function determined with the purely classical algorithm to the (also purely classically determined) optimal policy is 3.26.

- Given a random policy *beta*, in which an action is randomly chosen in each state, and a limited number of observations, we can use the QPU as a filter to identify the states or sequences of states that do not positively contribute to learning a policy which approximates an optimal policy. We can reduce the source state-action-value triples by up to 2/3, and by conducting a majority vote over the remaining actions per state we can approximate a policy that is equivalent or only slightly poorer than the policy found with the purely classical algorithm on all given state-action-value triples. By setting $n = 7$, $m = 500$ and running the algorithm 5 times and summarizing the result, we obtained 4904 state-action-value triples in the source data, with 4489 them being distinct. With the QPU-filter, we could reduce the number state-action-value triples to 1615. Due to the limited number of observations, we were not able to completely eliminate duplicate states with different actions in the quantum-classical algorithm. Thus, we applied a majority vote, and show that the results produced by the classical algorithm are only slightly better than the ones produced by the quantum-classical algorithm. However, the purely classical algorithm required 4904 state-action-value triples for making a decision, whereas the quantum-enhanced algorithm was able to filter out 3289 state-action-value triples and then find a qualitatively similar policy with only 1615 observations. In the described case, calculating the Euclidean distance from the policy found with the quantum-enhanced to the optimal policy found with the purely classical algorithm and 500,000 episodes (with varying numbers of states) is 10.2, whereas the distance from a policy determined with the purely classical algorithm to the (also purely classically determined) optimal policy is 9.8.

Summing up, we were able to directly embed maximally 7 complete state-value functions generated on 500 observations each on the QPU. With each execution the results my slightly vary, which is because of the randomly chosen observations, based on which different numbers of different states-action-value triples are available. We used between 1,100 and 1,700 qubits, a number which also varied with different sets of observations. The considered horizon $h$ can be of arbitrary length, and a discount factor may be used to scale the importance of the future states compared to the actual state, whereby we applied discount factors from 0.1 - 0.9 in our experiments. We were able to show that we can partially formulate policy-evaluation and iteration as QUBO, such that it can be presented to and solved with the support of a D-Wave 2000Q QPU. We were also able to show that by augmenting Monte Carlo policy evaluation, which calculates the value function for a given policy using sampling, with the introduced algorithm, we obtain equivalent or slightly better results compared to averaging over $n$ state-value functions or to each of the sub-optimal state-value functions (with identical $m$). The state-value functions we found based on the $n$ episodes and sub-optimal state-value functions are still not optimal, compared to classically found state-value functions based on tens or hundreds of thousands of observations. Figure 5 shows the state value function (assuming the player has a usable ace) based on the classical Monte Carlo algorithm. Figure 6 shows the state-value function based learned from averaging over 5 iterations and 7 classically found state-value functions. Figure 7 shows the results produced by the quantum-classical algorithm, which, in this case, produces a better than the classical algorithms with the same parameters.

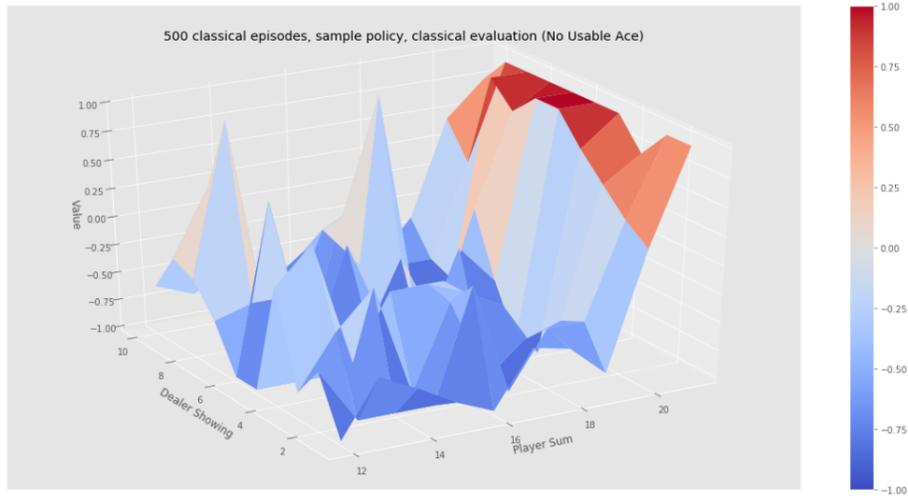

Figure 5: Classical Monte Carlo: 500 episodes, policy $\alpha$, no usable ace

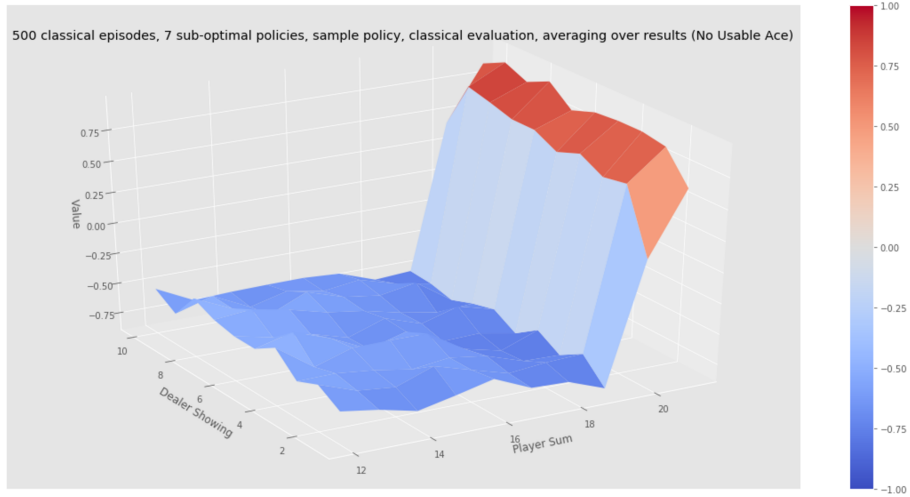

Figure 6: Classical Monte Carlo: 500 episodes, 7 policies, policy $\alpha$, no usable ace

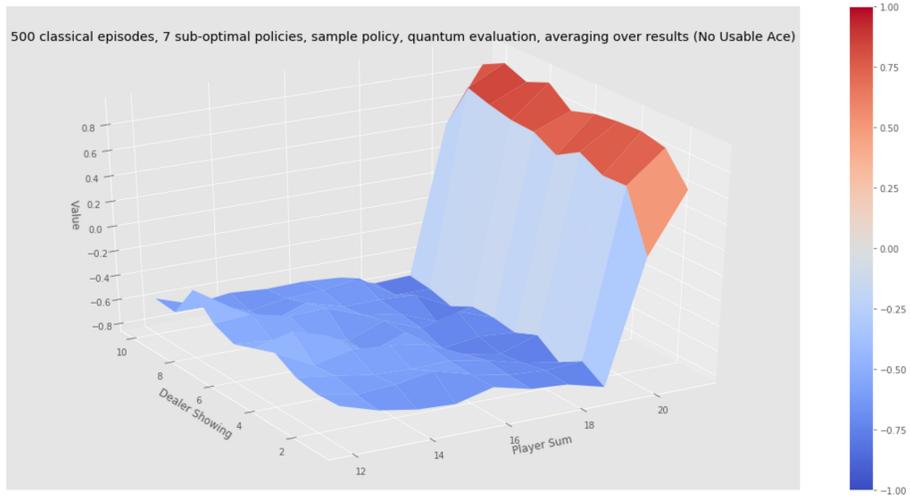

Figure 7: Quantum-enhanced Monte Carlo: 500 episodes, 7 policies, policy $\alpha$, no usable ace

# 7 Future work

In this report, we showed how to partially embed policy evaluation for discrete state spaces on the D-Wave QPU, whereas in our current work we aim to find quantum-classical algorithms capable of dealing with continuous state spaces, which is, i.e., ultimately useful in the context self-driving vehicles, where an agent needs to be able to make decisions considering a dynamically changing environment based on continuous state spaces. Furthermore, actions cannot necessarily be discretized, i.e., when we consider reinforcement learning in terms of self-learning/ healing machines, which we also aim to solve. We will continue to focus on solving practically relevant problems by means of quantum machine learning [22–27], quantum simulation, and quantum optimization.

# Acknowledgments

Thanks go to VW Group CIO Martin Hofmann and VW Group Region Americas CIO Abdallah Shanti, who enable our research.